# Method of measuring the chromatic dispersion parameter of optical fibers using a single-arm interferometer and a fiber femtosecond frequency comb

N.A. Koliada,[1,2,*] Y.G. Isaeva,[1,3] D.V. Brazhnikov,[1] A.A. Filonov,[1] and V.S. Pivtsov[1,3]

[1] Institute of Laser Physics SB RAS, Novosibirsk 630090, Russia
[2] Institute of Automation and Electrometry SB RAS, Novosibirsk 630090, Russia
[3] Novosibirsk State Technical University, Novosibirsk 630073, Russia

**ABSTRACT**. We propose and study a novel method for measuring the chromatic dispersion parameter ($D$) of optical fibers and bulk optical elements through the use of a single-arm three-wave interferometer (SAI) and a fiber femtosecond frequency comb (FFFC). The FFFC is frequency locked to a single-ytterbium-ion optical standard and employed as a source of highly stable and broadband laser radiation. The FFFC spectrum ranges from 1 to 2 μm, which is the most demanded range in fiber optics. The theoretical model presented in the work is used to derive analytical expressions for the parameter $D$, taking into account the second-order group velocity dispersion (GVD). To validate the methodology, the parameter $D$ is measured for a standard single-mode SMF-28 fiber with a length of 1.23 m. At a wavelength of 1550 nm, the value of $D$ has been found to be approximately 16.6 ps/(nm·km) with taking into account the second-order GVD. Concurrently, the root mean square error of measurements is only 0.86 ps/(nm·km), which is considerably less than the typical values for this type of fiber reported in the literature (5 ps/(nm·km)). Another important feature of the proposed method is that it enables the measurement of the parameter $D$ with high accuracy across a broad spectral range in a relatively short time.

## I. INTRODUCTION

The dispersion parameter $D$ of optical fibers is one of the most important parameters of optical fibers [1]. In order to design pulsed fiber laser systems, it is necessary to know $D$ with high accuracy, as it has a considerable impact on the temporal profile of the propagating light pulse. It must be accurately taken into account when designing complex fiber-optic systems, such as, for example, femtosecond fiber frequency combs [2, 3]. FFFCs have a wide range of applications [4, 5], including ultrafast laser spectroscopy [6–9], trace gas detection [10], calibration of astronomical spectrographs [11], high-harmonic generation [12], and high-accuracy laser detection and ranging (LADAR) [13, 14]. In quantum metrology, FFFC is one of the key elements of femtosecond optical clocks [15]. It is also actively used for comparing optical frequencies of different quantum frequency standards [16-20]. An accurate estimation of the dispersion parameter is also of significance for modern broadband telecommunication systems [21].

For many applied and fundamental research areas, it is extremely important to decrease the sensitivity of FFFCs to various external disturbances. This can be achieved by reducing the length of the fiber leads of the FFFC elements used, which can be done if their dispersion parameters are known with high accuracy. However, optical fiber manufacturers either provide insufficiently accurate $D$ values in datasheets or do not provide them at all. Moreover, the datasheets often do not include these data for non-standard dispersion-shifted, non-linear, or rare-earth doped optical fibers.

The three most common methods for determining the $D$ parameter are the phase shift method [22-24], the time-of-flight (spectral group delay) method [24-26], and the interferometric method [27-33]. The phase shift and time-of-flight methods are the most prevalent in commercial applications, as they allow dispersion measurement in optical fibers spanning distances from 1 km to several hundred km. The interferometric method is predominantly utilized in scientific investigations due to its capacity to perform dispersion measurements not only in long but also in short optical fibers, with lengths ranging from 0.01 m [24, 29, 34].

The advantage of the phase shift method is its fairly high measurement accuracy. The root mean square error (RMSE) for this method, as documented in reference [22], was 1 ps/(nm·km). However, the accuracy of this method is limited by the stability and frequency jitter of the radio frequency (RF) signal used. Also, ambient temperature fluctuations greatly affect the measurements. To reduce this effect, a series of measurements were conducted with and without thermal insulation of the optical fiber coil [26, 30].

---

*Contact author: n.koliada@mail.ru

A disadvantage of the time-of-flight method is the influence of the shape of the laser pulses used on their temporal broadening, which leads to errors in measuring *D*. Furthermore, in order to achieve a RMSE of 1.5 ps/(nm·km), the fiber length should be a minimum of 7.8 m [35]. This means that this method is unsuitable for the design of precision laser systems utilizing short optical fibers.

The measurement accuracy of both phase shift and time-of-flight methods depends on the length of the optical fiber. To achieve high measurement accuracy, the sample length should be in the range of tens of meters to several hundred kilometers. Furthermore, a reference fiber with a length of no more than one meter is required, against which subsequent calculations will be performed [22-26]. Moreover, relatively complex mathematical transformations are required to process the data [28].

There are two categories of interferometric methods: temporal and spectral. The RMSE of the measured dispersion using temporal interferometry in [36] was 1.8 ps/(nm·km), when examining a photonic crystal fiber with a length of 0.814 m. The main disadvantage of time interferometry is that it is sensitive to external disturbances due to the mobility of the air arm of the interferometer, which leads to an increase in the RMSE.

Similar to temporal interferometry, spectral interferometry is also capable of characterizing dispersion in short fibers, with a length of less than one meter. Spectral interferometry is typically more stable than temporal interferometry. Moreover, this method is quite simple to implement. In spectral interferometry, two-arm interferometers usually have a Mach-Zehnder [34, 37] or Michelson [38, 39] configuration. The Michelson configuration is the most commonly utilized configuration. One of the disadvantages of the fiber-based dual-arm interferometer configuration is that the lengths of the fiber leads of the splitter must be identical. Otherwise, an alternative set of interference fringes is generated, resulting in the distortion of the spectral interferogram and an increase in the RMSE. Moreover, in the case of pulsed radiation, precise matching of the fiber lead lengths is necessary to combine the pulses in time. In addition, the radiation experiences different temperature and mechanical external influences in each individual fiber channel (due to the sensitivity of the fiber characteristics to these disturbances), which makes the interferogram less contrasting.

An alternative configuration is the single-arm interferometer [29]. Interferograms obtained with the single-arm and two-arm interferometers are found to be equivalent to each other. Nevertheless, the RMSE of the dispersion parameter recorded with the single-arm interferometer is 0.25 ps/(nm·km) (in a 0.395 m long single-mode fiber) [29], whereas that obtained with the two-arm interferometer is 0.7 ps/(nm·km) (1 m sample length) [28]. The RMSE of the single-arm interferometer is smaller due to its lower sensitivity to external perturbations. Furthermore, such an interferometer is easier to implement. Moreover, the value of *D* can be calculated from experimental data through the application of straightforward mathematical formulas. An additional advantage of this method is that the amplitudes of the interfering fields are not involved in the calculation of the dispersion parameter (as will be shown below). This simplifies the implementation and use of the interferometer.

In [28, 29], an SAI scheme for measuring the parameter *D* was investigated with a tunable diode laser serving as the laser radiation source. The disadvantage of this approach is the short coherence length, which means that *D* can only be measured for fibers less than 1 m long. In addition, the wavelength tuning is performed in a narrow range of 1550±65 nm, which constrains the measurement range of the *D* parameter. Moreover, these works do not provide a complete theoretical model of the SAI. Specifically, they do not provide explicit analytical expressions for calculating *D* and do not consider the influence of second-order GVD, which can result in a notable correction to the *D* value.

Here, we propose using a fiber femtosecond frequency comb as a source of laser radiation for the SAI instead of a tunable diode laser. The multimode radiation of the FFFC is locked to the optical frequency standard based on a single ytterbium ion [15, 40, 41]. The emission spectrum of the FFFC covers a range of wavelengths from 1 to 2 μm. Therefore, it is possible to measure the dispersion parameter within this broad range. The frequency-stabilized FFFC light exhibits high coherence, meaning that the dispersion of both short-length (less than a meter) and long-length (kilometers) fibers can be measured. Another advantage of the proposed method lies in its operational speed, i.e. a short time is required to acquire an interferogram. Indeed, an interferogram is recorded almost instantaneously using an optical spectrum analyzer over a wide range of wavelengths, i.e. no time is wasted for reconfiguring the wavelength of the laser source and recording the spectrum at each wavelength. The high coherence of the laser radiation used results in high contrast of the interferograms, which facilitates the accurate acquisition of data for calculating the dispersion parameter.

To describe the operation of the SAI/FFFC, we have developed a theoretical model and, for the first time, obtained explicit analytical expressions for the $D$ parameter, taking into account the dispersion of high-order group velocities. These expressions have allowed us to distinguish and estimate the influence of the second-order GVD. The performance of the proposed method has been experimentally validated using a 1.23 meter long SMF-28 test fiber.

## II. EXPERIMENTAL SETUP

Fig. 1 shows the SAI scheme for measuring the dispersion parameter $D$. The system is a two-pass system.

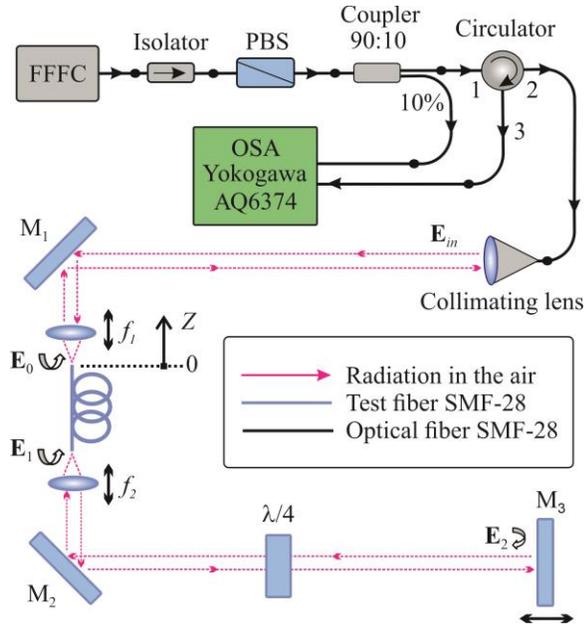

FIG 1. Scheme of a single-arm three-wave interferometer for measuring the dispersion parameter $D$. FFFC – fiber femtosecond frequency comb [2]; PBS – polarizing beam splitter with fiber outputs; $M_1$, $M_2$, $M_3$ – flat metal mirrors; $\lambda/4$ – quarter-wave phase plate; $E_0$, $E_1$ and $E_2$ – the waves contributing to the interference pattern, while $E_{in}$ is an incident wave.

The FFFC with an intermode frequency of ~82.6 MHz [22, 23], developed at the Institute of Laser Physics SB RAS, was used as a source of laser radiation. It was stabilized by the frequency of the optical standard on a single ytterbium ion [24]. Fig. 2 shows the optical spectrum at the input of the interferometer (derived from the 10% fiber coupler output). The FFFC is a device that generates a femtosecond optical frequency comb. When the FFFC is stabilized by the frequency of an optical standard, each component of the comb acquires the stability of that standard. At the same time, femtosecond pulses are observed in the time domain.

The interference pattern observed at the output of the single-arm interferometer is the result of the superposition of three waves: two reflected from the facets of the fiber under study and one from the mirror $M_3$ (Fig.1).

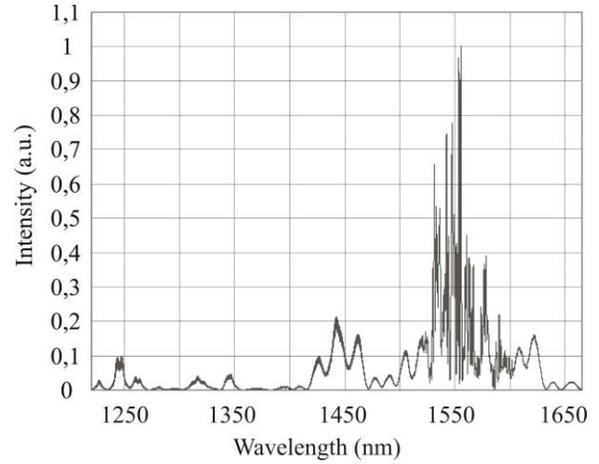

FIG 2. Optical spectrum at the input of the single-arm interferometer (a part of the optical spectrum of the FFFC). The resolution is 0.05 nm.

A 90:10 broadband fiber coupler diverts 10% of the laser power to the Yokogawa AQ6374 Optical Spectrum Analyzer to record the optical spectrum at the interferometer input. 90% of the radiation enters a fiber-optic circulator with FC/APC-type connectors. A fiber-optic collimator with an anti-reflective coating for wavelengths of 1050-1650 nm is used to collimate the laser radiation and its subsequent input into the test fiber. The SMF-28 fiber segment was chosen as the test fiber because its characteristics are known and can be compared with the values obtained during the work. Flat metal mirrors ($M_1$, $M_2$ and $M_3$) are used to tune the interferometer and to inject radiation back into the test fiber. Lenses ($f_1$ and $f_2$) with focal lengths of 8 mm are used to focus the radiation into the test fiber and to collimate the radiation at the fiber exit. The lenses are mounted on miniature linear translation stages with microscrews that allow them to be moved with high precision along the axis of radiation propagation. The $M_3$ mirror is also mounted on a linear translation stage with a range of motion of 13 mm, which allows tuning of the balanced wavelength $\lambda_0$. The balanced wavelength is the wavelength of laser radiation at which the optical lengths of the test fiber and the air path coincide. In this case, the interferometer is called balanced (see section 3). The quarter-wave plate allows to adjust the contrast of the interference fringes by matching the polarization parameters of the interfering waves. The length of the test fiber is 1.23 m, the length of the air path is 1.8 m. The lengths were chosen such that femtosecond laser pulses reflected from different surfaces, i.e., waves $\mathbf{E}_0$, $\mathbf{E}_1$ and $\mathbf{E}_2$ coincide in time at the point $z = 0$ and interference occurs. Additional reflections that could occur at the facets of the fiber elements are eliminated by using FC/APC connectors.

Fig. 3 shows the interference pattern obtained experimentally in the proposed scheme (Fig. 1). In this form, the interferogram is difficult to analyze because the spectrum is not uniform in amplitude and the result of interference is not visible. Therefore, the obtained spectrum was normalized to the spectrum at the input of the interferometer. The resulting spectrum is shown in Figure 4, where $\lambda_0$ is the balanced wavelength (see Sec. 3), while $\lambda_1$, $\lambda_2$, $\lambda_{-1}$, $\lambda_{-2}$ are the wavelengths of interference peaks on both sides of the balanced wavelength.

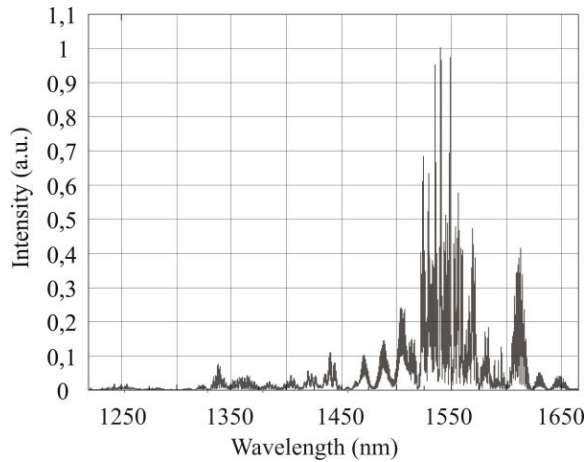

FIG 3. Optical spectrum at the output of a single-arm interferometer (a three-wave interferogram). The resolution is 0.05 nm.

Fig. 5 shows a narrow section of the interferogram illustrated in Fig. 4. It can be seen that there is no additional high-frequency envelope of the signal. Consequently, no supplementary parasitic interference occurs in the interferometer other than that which leads to the modulation of the envelope shown in Fig. 4.

To further analyze the experimental data and calculate the dispersion parameter $D$, a theoretical model of SAI was developed and presented in the subsequent chapter.

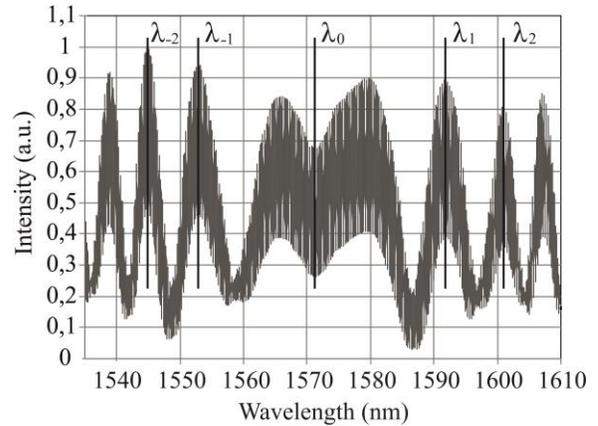

FIG 4. Experimentally obtained interference pattern with balanced wavelength around 1615 nm in the range of 1585-1645 nm. The resolution is 0.05 nm.

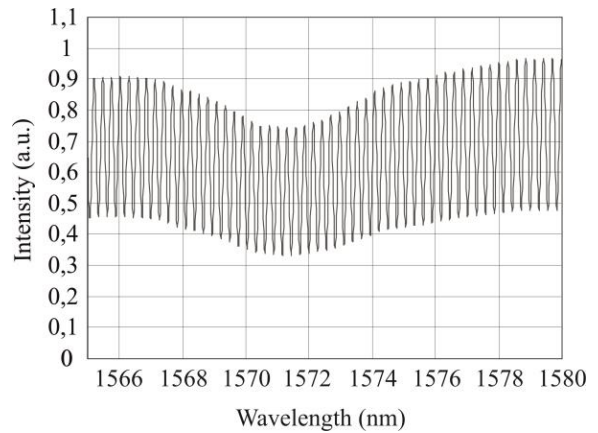

FIG 5. Zoom of the interferogram depicted in Fig. 4.

### III. THEORY

In the single-arm interferometer (Fig. 1), the chromatic dispersion parameter $D$ is calculated using the envelope of the interference pattern (Fig. 4), which is primarily influenced by three waves: $\mathbf{E}_0$, $\mathbf{E}_1$ and $\mathbf{E}_2$. We neglect other contributions associated with multiple reflections of light from optical elements due to their small influence. The origin of the $z$-axis, along which the waves propagate, is aligned with the input face of the fiber (Fig. 1). Then the three waves indicated above and the incident wave from the source ($\mathbf{E}_{in}$) at a point $z = 0$ can be written as:

$$\mathbf{E}_{in}(t) = E_{in}\mathbf{e}_{in}e^{-i\omega t}, \quad (1)$$

$$\mathbf{E}_0(t) = E_0\mathbf{e}_0 e^{-i\omega t}, \quad (2)$$

$$\mathbf{E}_1(t) = E_1\mathbf{e}_1 e^{-i\omega t}, \quad (3)$$

$$\mathbf{E}_2(t) = E_2\mathbf{e}_2 e^{-i\omega t}. \quad (4)$$

Here $\mathbf{e}_{in}$, $\mathbf{e}_{0,1,2}$ are unit complex vectors describing the polarization of light, which is generally elliptical. The amplitudes of the interfering waves $E_0$, $E_1$ and $E_2$ can be expressed in terms of the amplitude of the incident wave $E_{in}$ using the Fresnel coefficients for normal incidence, which, in this case, are real and positive numbers.

$$E_0 = -\alpha E_{in}, \; E_1 = \beta E_{in}, \; E_2 = -\gamma E_{in}, \quad (5)$$

where the sign "−" before $\alpha$ and $\gamma$ is used to indicate the phase jump of the wave when it is reflected from a material with greater optical density. It is evident that the condition is satisfied for the introduced coefficients: $\alpha, \beta, \gamma < 1$.

The intensity of the total light field at $z = 0$ is determined by the following expression:

$$\begin{aligned} I &= \frac{c}{2\pi}\left|\mathbf{E}_0 + \mathbf{E}_1 + \mathbf{E}_2\right|^2 \\ &= \alpha^2 + \beta^2 + \gamma^2 - 2\alpha\beta\,\mathrm{Re}\left\{\mathbf{e}_0\mathbf{e}_1^* e^{-2ik_1L_1}\right\} \\ &\quad + 2\alpha\gamma\,\mathrm{Re}\left\{\mathbf{e}_0\mathbf{e}_2^* e^{-2i(k_1L_1+k_2L_2)}\right\} \\ &\quad - 2\beta\gamma\,\mathrm{Re}\left\{\mathbf{e}_1\mathbf{e}_2^* e^{-2ik_2L_2}\right\}. \end{aligned} \quad (6)$$

Here $L_1$ and $L_2$ are the length of the test fiber and the length of the air path, respectively.

To simplify the theory, we assume that the residual ellipticity of polarization that can be acquired by linearly polarized light when passing through an optical fiber is minimal and can be neglected. Therefore, in the theoretical model we are considering, all interfering fields have linear polarization. If we combine the $x$-axis with the direction of linear polarization of the field $\mathbf{E}_0$, then in the Cartesian basis $\mathbf{e}_x = \{1,0\}$, $\mathbf{e}_y = \{0,1\}$ the polarization vectors of the interfering fields have the following coordinates:

$$\mathbf{e}_0 = \{1,0\}, \; \mathbf{e}_1 = \{\cos\varphi_1, \sin\varphi_1\},$$
$$\mathbf{e}_2 = \{\cos\varphi_2, \sin\varphi_2\}. \quad (7)$$

where $\varphi_m$ is the angle between the linear polarization of the $\mathbf{E}_m$ field and the $x$-axis ($m = 1,2$).

The reflection from the fiber facets is relatively small, thereby satisfying the condition: $\alpha, \beta \ll \gamma$. Taking into account this condition and (7), expression (6) takes the form:

$$I \approx \gamma^2 + 2\alpha\gamma\cos\varphi_2\cos(\xi_1 + \xi_2)$$
$$- 2\beta\gamma\cos(\varphi_1 - \varphi_2)\cos\xi_2. \quad (8)$$

where the notation for the spatial phase overlap is introduced:

$$\xi_m = 2k_m L_m. \quad (9)$$

As can be seen from (8), the interference pattern is the sum of two cosines, each of which contributes proportionally to the amplitudes $2\alpha\gamma\cos\varphi_2$ and $2\beta\gamma\cos(\varphi_1 - \varphi_2)$. At the same time, expression (8) does not describe the usual beats between two harmonic signals, since the arguments of the cosines are not close to each other. This represents a key distinction between our theoretical model and that presented in [11].

By using an auxiliary quarter-wave plate ($\lambda/4$) in the scheme of the proposed single-arm interferometer, the rotation angle $\varphi_2$ can be tuned. This approach allows for the optimization of the contrast of the interference pattern, thereby enhancing the precision with which the position of the envelope peaks in Fig. 4 can be determined. However, since the amplitudes of the interference peaks themselves are not employed in the calculation of the dispersion parameter, we can disregard the dependence of expression (8) on the polarization angles of the interfering fields, thereby simplifying the analysis. In other words, it can be assumed that all of the waves involved in the interference are polarized along the $x$-axis. This ultimately yields the final expression that describes the interference pattern:

$$I \approx \gamma^2 + 2\alpha\gamma\cos(\xi_1 + \xi_2) - 2\beta\gamma\cos\xi_2. \quad (10)$$

It follows from this expression that the maxima of the slow envelope of the interference pattern will occur when two conditions are simultaneously fulfilled:

$$\begin{cases} \xi_1 + \xi_2 = 2\pi q_1, \\ \xi_2 = (1 + 2q_2)\pi, \end{cases} \quad (11)$$

where $q_1$, $q_2 = 0, \pm 1, \pm 2$. Using (9), the conditions (11) can be rewritten in another form:

$$\begin{cases} 2(nL_1 + L_2) = q_1 \lambda, \\ 2L_2 = (1 + 2q_2)\dfrac{\lambda}{2}. \end{cases} \quad (12)$$

The first expression in (12) means that the doubled optical path from the beginning of the test fiber to the mirror must include an integer number of wavelengths. The second condition stated in (11) requires an odd number of half-wavelengths within the doubled air gap length.

Expressions (1) – (4) describe monochromatic fields. In the experiment, a laser source of femtosecond pulses is used (see Fig. 6). In this instance, the three-wave interference will occur if the pulses corresponding to waves $\mathbf{E}_0$, $\mathbf{E}_1$ and $\mathbf{E}_2$ are combined in space after passing through a single-arm interferometer (at $z = 0$).

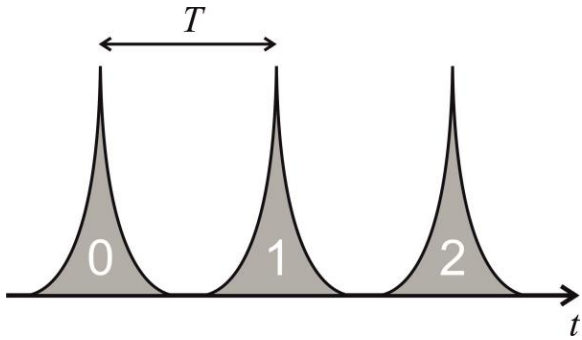

FIG 6. Conditional image of light pulses at the interferometer input, following each other with period $T$.

Consider the moment in time when the third pulse of light arrives at the input face of the fiber. By this time, the second pulse must have traveled back and forth across the test fiber to interfere with the third pulse. This occurs when the condition is met.

$$\frac{2L_1}{\upsilon_g} = T. \quad (13)$$

Here $\upsilon_g$ is the group velocity. Similarly, the first pulse must travel both ends of the fiber and the air gap in time $2T$, i.e.

$$\frac{2L_1}{\upsilon_g} + \frac{2L_2}{c} = 2T, \quad (14)$$

where $c$ is the speed of light in vacuum. Combining conditions (13), (14), we come to the following requirement:

$$n_g L_1 = L_2, \quad (15)$$

where $n_g = c/\upsilon_g$ is the group refractive index of the fiber, which is related to the refractive index $n$ in a known way: $n_g = n - \lambda n^{(1)}$. Hereinafter, the upper indices $n$ of the species ($j$) will mean the derivative of $j$-order by $\lambda$: $n^{(j)} \equiv d^j n / d\lambda^j$. Thus, in order to realize three-wave interference, the condition must be satisfied:

$$\left(n - \lambda n^{(1)}\right) L_1 - L_2 = 0. \quad (16)$$

In the experiment, the position of the mirror $M_3$ is adjusted so that condition (16) is satisfied. At the same time, since $\lambda$ enters (16) explicitly, and because of the presence of fiber dispersion ($n^{(1)} \neq 0$), condition (16) will be satisfied only for a certain radiation wavelength, which, as in [29], we will call balanced and denote $\lambda_0$.

To engage condition (16), consider the difference of the two conditions from (11):

$$\xi_1 - \xi_2 = 2\pi q, \quad (17)$$

where $q = 0, \pm 1, \pm 2$. Using (9), this expression can be rewritten as:

$$\frac{2}{\lambda}(L_1 n - L_2) = q. \quad (18)$$

Since $n$ in the considered region is a smooth function of $\lambda$, we can use the approximate Taylor series expansion near $\lambda=\lambda_0$:

$$n(\lambda) \approx n_0 + n_0^{(1)}(\lambda - \lambda_0)$$
$$+ \frac{n_0^{(2)}}{2}(\lambda - \lambda_0)^2 + \frac{n_0^{(3)}}{6}(\lambda - \lambda_0)^3. \quad (19)$$

Here, the series expansion is performed up to the third order to take into account the influence of the second-order GVD associated with $n^{(3)}$. For convenience, notations have been introduced in (19):

$$n_0 = n(\lambda_0), \quad n_0^{(j)} = \left.\frac{d^j n}{d\lambda^j}\right|_{\lambda=\lambda_0}. \quad (20)$$

Substituting (19) into (18), we have:

$$\frac{2}{\lambda}\left(L_1 n_0 - L_1 \lambda_0 n_0^{(1)} - L_2\right)$$
$$+ \frac{2}{\lambda} L_1 \left(\lambda n_0^{(1)} + \frac{1}{2} n_0^{(2)}(\lambda - \lambda_0)^2\right.$$
$$\left. + \frac{1}{6} n_0^{(3)}(\lambda - \lambda_0)^3\right) = q. \quad (21)$$

Considering that condition (16) is satisfied for $\lambda_0$, the first bracket in (21) turns to zero. Then we finally arrive at the expression, which is satisfied by the maxima of the envelope of the interference pattern:

$$L_1\left(2n_0^{(1)} + n_0^{(2)}\frac{(\lambda - \lambda_0)^2}{\lambda} + n_0^{(3)}\frac{(\lambda - \lambda_0)^3}{3\lambda}\right) = q. \quad (22)$$

It should be noted that, by analogy with the derivation of expression (22), an expression can be derived that is satisfied by the minima (nodes) of the envelope of the interference pattern. To calculate the parameter $D$ taking into account the second-order GVD, we use both the distance between the peaks $\lambda_2$ and $\lambda_1$, and between the peaks, $\lambda_{-1}$ and $\lambda_{-2}$ (see Fig. 4). In general, the value of $q$ in (22) can be any integer, including zero. Nevertheless, the increment ($\Delta q$) between neighboring peaks is always +1 or −1. It can be shown that the sign is contingent upon the sign of the derivative $n_0^{(1)}$. Thus, for the first pair of peaks situated to the right of $\lambda_0$ in Fig. 4, the result obtained from (22) is:

$$q_2 - q_1 = \text{sign}\left(n_0^{(1)}\right)$$
$$= L_1\left(2n_0^{(1)} + n_0^{(2)}\frac{(\lambda_2 - \lambda_0)^2}{\lambda_2} + n_0^{(3)}\frac{(\lambda_2 - \lambda_0)^3}{3\lambda_2}\right)$$
$$- L_1\left(2n_0^{(1)} + n_0^{(2)}\frac{(\lambda_1 - \lambda_0)^2}{\lambda_1} + n_0^{(3)}\frac{(\lambda_1 - \lambda_0)^3}{3\lambda_1}\right)$$
$$= \frac{L_1 \Delta\lambda_{21}}{3}\left\{3\left[1 - l_1\right]n_0^{(2)}\right.$$
$$\left. + \left[\lambda_{12} - \lambda_0(3 - l_1)\right]n_0^{(3)}\right\}. \quad (23)$$

Here for convenience we have introduced the notations: $\Delta\lambda_{ij}=\lambda_i-\lambda_j$, $\lambda_{ij}=\lambda_i+\lambda_j$, $l_1 = \lambda_0^2/\lambda_1\lambda_2$. Note that for the considered fiber $\text{sign}\left(n_0^{(1)}\right) = -1$. Indeed, given that in the case of $\text{sign}\left(n_0^{(1)}\right) = -1$ the refractive index increases with decreasing $\lambda$, the interference peaks follow each other more frequently. This phenomenon is exemplified by the fiber we utilize, as seen from Fig. 4 where $\Delta\lambda_{-1-2} < \Delta\lambda_{21}$.

Similarly, for the distance between the peaks $\lambda_{-1}$ and $\lambda_{-2}$:

$$q_{-2} - q_{-1} = \text{sign}\left(n_0^{(1)}\right) = \frac{L_1 \Delta\lambda_{-2-1}}{3}\left\{3\left[1 - l_2\right]n_0^{(2)}\right.$$
$$\left. + \left[\lambda_{-1-2} - \lambda_0(3 - l_2)\right]n_0^{(3)}\right\}, \quad (24)$$

with $l_2 = \lambda_0^2/\lambda_{-1}\lambda_{-2}$.

Equations (23), (24) form a system in which the unknowns are $n_0^{(2)}$ and $n_0^{(3)}$. Since for the calculation of $D$ we need only the solution for $n_0^{(2)}$, solving the system (23), (24), we find:

$$n_0^{(2)} = \tilde{n}_0^{(2)} \times \frac{1 + \eta}{1 + \frac{\Delta\lambda_{-2-1}(1 - l_2)}{\Delta\lambda_{21}(1 - l_1)} \times \eta}. \quad (25)$$

The following notations are introduced here:

$$\tilde{n}_0^{(2)} = \frac{\text{sign}\left(n_0^{(1)}\right)}{L_1 \Delta\lambda_{21}(1 - l_1)}, \quad (26)$$

$$\eta = \frac{\Delta\lambda_{21}\left[\lambda_{12} - \lambda_0\left(3 - l_1\right)\right]}{\Delta\lambda_{-1-2}\left[\lambda_{-1-2} - \lambda_0\left(3 - l_2\right)\right]}. \quad (27)$$

At that, if we do not take into account the influence of the second-order GVD ($n_0^{(3)} = 0$), we should put $\eta = 0$ and $n_0^{(2)}$ go to $\tilde{n}_0^{(2)}$.

The specific chromatic dispersion parameter is defined by the following expression [1]:

$$D = -\frac{\lambda_0}{c} n_0^{(2)}. \quad (28)$$

Then from (25) we arrive at the final expression:

$$D = -\frac{\lambda_0}{c} \tilde{n}_0^{(2)} \times \frac{1 + \eta}{1 + \frac{\Delta\lambda_{-2-1}\left(1 - l_2\right)}{\Delta\lambda_{21}\left(1 - l_1\right)} \times \eta}. \quad (29)$$

### IV. RESULTS AND DISCUSSION

During the work, interference patterns were recorded for the SMF-28 test fiber at balanced wavelengths within the range of 1430-1630 nm. Fig. 7 illustrates the relationship between the dispersion parameter $D$ and the wavelength of laser radiation. The red line represents the dispersion parameter without considering the second-order GVD, while the green line incorporates the second-order GVD, calculated using expression (29).

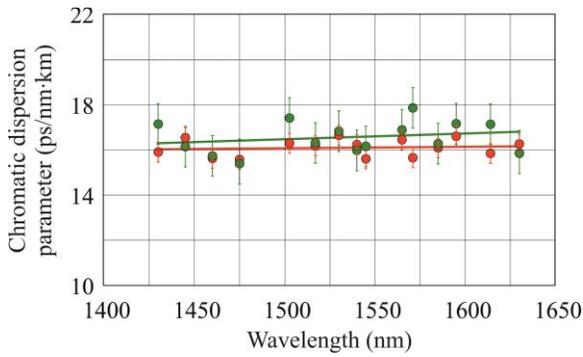

FIG 7. The spectral dependence of the dispersion parameter $D$ for the SMF-28 test fiber without taking into account (red line) and taking into account (green line) the second order of GVD.

It is generally accepted that the second-order GVD makes a noticeable contribution when the first-order GVD is close to zero [1]. However, Fig. 7 shows that the spectral dependence of the parameter $D$ taking into account the second-order GVD has larger values and a larger slope angle in the region of the studied wavelengths. Therefore, the second-order GVD should be taken into account when designing fiber laser systems.

In particular, the parameter $D$ at a wavelength of 1550 nm was 16.1 ps/(nm·km) without taking into account the second-order GVD and 16.6 ps/(nm·km) with taking into account the second-order GVD. The RMSE was 0.42 ps/(nm·km) and 0.86 ps/(nm·km), respectively. The standard value of the $D$ parameter given in various commercial sources for the SMF-28 fiber for an emission wavelength of 1550 nm is 18 ps/(nm·km), and the RMSE is 5 ps/(nm·km). Thus, the RMSE values obtained in our work are noticeably smaller than the standard ones

In the current conditions, the spectral range of dispersion parameter measurements is constrained by the transmission spectrum of fiber elements within the interferometer scheme. The substitution of these elements with those of a broader bandwidth will facilitate a considerable expansion of the measurement spectral range in the future.

### V. CONCLUSIONS

The work reports on the experimental realization of a single-arm three-wave interferometer for measuring the dispersion parameter of optical fibers. For the first time, a fiber femtosecond frequency comb phase-locked to the optical single-ion frequency standard is employed as a source of laser radiation for this type of interferometer. The extremely high coherence of the FFFC radiation results in high-contrast and stable interferograms that are recorded almost instantaneously across a wide wavelength range, unlike other schemes, using tunable single-frequency lasers.

We have developed a theoretical model, which allowed us to derive an explicit analytical expression for the dispersion parameter. This expression, for the first time for this type of interferometer, reflects the influence of the second-order group velocity dispersion, which is of interest for a wide range of applications. The high accuracy of the experimental measurements demonstrates that the second-order GVD makes a notable contribution even in circumstances where the first-order GVD is significantly different from zero, despite previous assumptions to the contrary. Another advantage of the proposed method is that it can be efficiently employed for both short (< 1 m) and long (>> 1 m) optical fibers.

The proposed scheme has the potential to further reduce the RMSE of the measurements. Furthermore, it permits the consideration of GVD of higher orders and the measurement of the dispersion parameter of non-standard optical fibers by substituting the test fiber with them, as well as short fibers with lengths ranging from 1 cm and various bulk optical elements (crystals, glasses, chirped mirrors, etc.) by positioning them in the air path. Due to high accuracy of the propose method, it paves the way for designing fiber based and other types of optical metrology systems with minimum length and low sensitivity to external perturbations.

**ACKNOWLEDGMENTS**

This work was partially supported by the state assignment of Institute of Laser Physics SB RAS (Project numbers 121033100064-9 and 121041300256-1). The equipment of the Collective Use Center "Femtosecond Laser Complex" was involved during the work. This work was also partially supported by the state assignment of Institute of Automatics and Electrometry SB RAS (project number FWNG-2024-0015). We express our gratitude to E.V. Podivilov from Institute of Automatics and Electrometry SB RAS for useful discussions.